%% file: aanda.tex
\documentclass{aa}  

\usepackage{graphicx}
\usepackage{txfonts}
\usepackage{mathrsfs}
\usepackage{stfloats} 
\usepackage[usenames,dvipsnames]{color}
\usepackage[normalem]{ulem}
\usepackage{enumitem}
\usepackage{url}
\usepackage[colorlinks=true,
    linkcolor=blue,
    filecolor=magenta,      
    urlcolor=blue,
    citecolor=blue]{hyperref}

\usepackage{subcaption}

\newcommand*{\Msun}{M$_\odot\ $}

\newcommand{\firstrev}[1]{ #1}

\begin{document}

   \title{Massive Stellar Cannibals: How Stellar Mergers Drive Mass-Loss in Extremely Massive Stars}


   \author{J. Roman-Garza \inst{1,2} , 
   T. Fragos \inst{1,2},
   C. Charbonnel \inst{1,2,3},
   L. Ramírez-Galeano \inst{1},
   M.\,Kruckow \inst{1,2},
   E. Farag \inst{4}
   }

   \institute{Département d’Astronomie, Université de Genève, Chemin Pegasi 51, CH-1290 Versoix, Switzerland 
         \and
         Gravitational Wave Science Center (GWSC), Université de Genève, CH1211 Geneva, Switzerland. 
         \and
         IRAP, UMR 5277 CNRS and Université de Toulouse, 14 Av. E.Belin, 31400 Toulouse, France \and
          Department of Astronomy, Yale University, New Haven, CT 06511, USA\\
             \email{jaime.romangarza@unige.ch}}

   \date{Received xxx; accepted xxx}

 \authorrunning{J. Roman-Garza et al.} \titlerunning{Massive Stellar Cannibals: How Stellar Mergers Drive Mass-Loss in Extremely Massive Stars}
 
  \abstract
   {It has been theorized that the formation of extremely massive and supermassive stars ($>10^3\ {\rm M}_\odot$) could plausibly be the outcome of stellar mergers in low metallicity ($Z<10^{-1}$~Z$_\odot$) and dense ($\gtrsim10^3\ {\rm M}_\odot\ {\rm pc}^{-3}$) stellar environments. These objects remain relevant as they can serve as the progenitors of intermediate-mass black holes and they are also formidable chemical polluter candidates, as evidenced by the peculiar abundances seen across cosmic history. Our understanding of the formation of these objects depends on the physical processes involved in their mass accretion and loss,   via gas inflows and stellar winds, as well as the stellar mergers themselves. 
   } 
   {This work investigates merger-induced mass loss in extremely massive stars within a hydrodynamic framework and provides a prescription derived from the simulations to estimate both the mass loss and the outcome of the interaction. 
   }
   {We  adapted the 1D hydrodynamic, stellar structure, and evolution code {\tt MESA} to simulate stellar inspirals. In our simulations, we considered stars of $>1000\,\rm M_{\odot}$ with inspiraling companions  of $<100$ M$_\odot$; hence, with mass ratios of $<0.1$. As the inspiral progresses, the orbital energy of the system is lost through the hydrodynamic and gravitational drag forces. This energy gets deposited as thermal energy in the extremely massive star's envelope. The reaction of the star can be followed by solving Euler's hydrodynamical equations.  
   }
   {The extremely massive star experiences mass loss from pulsations produced by the inspiral. By scaling the mass lost through such pulsations based on the time it takes for the extremely massive star to radiate away the injected thermal energy, we find that the total ejected mass is $\sim$10-30$\%$ of the system's mass. 
   Our results point out that most of the energy deposited by the inspiral is used to eject mass. This is consistent with the fact that the extremely massive star models that we study are barely bound, as their mean adiabatic exponent is $\sim 4/3$.
   }
   {These findings demonstrate that merger-induced mass loss is non-negligible for the considered configurations. Thus, it is  an important process to account for when investigating the formation of extremely massive stars and predicting their possible role throughout cosmic history.
   }

   \keywords{stars: massive, stars: mass-loss 
               }

   \maketitle

%

\section{Introduction}

In dense and massive stellar clusters the constituents of the system segregate by mass. As the central density rises until the collapse of the cluster \citep[][]{bettwieser1984post} an increase in close gravitational interactions between stars is experienced, possibly leading to the coalescence of these stars \citep{hills1976stellar,portegies1999star}. The dynamical merger of stars is considered a scenario to create peculiar objects such as blue stragglers \citep[see][and references there in]{mathieu2025blue}, as well as very massive stellar objects (VMSOs). The stellar mass spectrum \citep[following the definition by][]{gieles2025globular} given by 
\begin{itemize}
    \item very massive stars (VMSs), with $10^2$ to $<10^3$ M$_\odot$;
    \item extremely massive stars (EMSs), with $10^3$ to $<10^4$ M$_\odot$; 
    \item supermassive stars (SMSs), with masses equal or above $10^4$ M$_\odot$.
\end{itemize}

Despite the lack of direct observational evidence for the existence of EMSs and SMSs, considering their presence across the cosmic history is relevant as they are formidable candidates for: intermediate mass black hole (IMBH) progenitors \citep[][]{portegies2004formation,reinoso2018collisions,volonteri2010formation,devecchi2009formation}; chemical polluters for proto-globular clusters, evidenced by peculiar chemical abundances in their stellar populations \citep{denissenkov2013supermassive,gieles2018concurrent,gieles2025globular,ramirezga2025}; and explaining the enhanced nitrogen abundances observed in extremely compact star-forming galaxies observed in the early universe \citep[see][]{charbonnel2023n,marques2024extreme,2024A&A...687L..11S} as well as the observed features of the compact emitters so called as "little red dots" \citep[e.g.][]{begelman2025little,nandal2025supermassive,coughlin2024quasi} .

Nevertheless, a complete self-consistent theory for the dynamical formation of VMSOs requires understanding the role of mass accretion and loss processes experienced in dense stellar environments. The main relevant processes are: stellar mergers contributing to the mass accretion and loss \citep[e.g.][]{glebbeek2009evolution, glebbeek2013structure,gaburov2010onset,ramirezga2025, gieles2018concurrent}; as well as the contribution of mass by gas accretion \citep[e.g.][]{gieles2025globular,gieles2018concurrent} and stellar winds driving mass-loss \citep[see][]{ramirezga2025,gieles2025globular,glebbeek2009evolution,2025A&A...699A..71H}. 

In particular, collision induced mass-loss has been investigated mainly in a hydrodynamical framework for lower mass stars, i.e. from 1 to $\sim50$ M$_\odot$, in the context of triple systems \citep[see][]{gaburov2010onset} and involved in head-on collisions \citep[see][]{glebbeek2013structure}; there it is found that a maximum loss of $10\%$ to $25\%$ of the system mass results from these interactions. Those results depend on the configurations involved and the structure of each star.  \cite{ramirezga2025} provides a semi-analytical prescription to determine the mass-loss induced by collisions for VMSOs created by high cold-gas accretion rates, while no hydrodynamical framework has been used to provide answers for this range of stellar masses. 

This work aims to study the resulting mass-loss from stellar mergers involving EMSs and less massive stars. Approaching this problem in a complete and self-consistent manner is beyond our capabilities, as this would require 3D hydrodynamic simulations with radiation transport. Due to the wide range of temporal and spatial scales, this task becomes very challenging, if not impossible. Given such limitations, we use a 1D framework \citep[adapted from the one used in][]{fragos2019complete} built over the Modules for Experiment in Stellar Astrophysics (\texttt{MESA}) stellar evolution software instrument \citep{mesa1,mesa2,mesa3,mesa4,mesa5,mesa6} and taking advantage of its Riemann hydrodynamic solver to model the response of the EMS structure during stellar circular inspirals. We present how the information gathered from circular inspirals can be used to determine the interaction outcomes across initial orbital configurations. The general methodology on the 1D hydrodynamic and stellar evolution framework used is discussed in Sec.~\ref{sec:methods_col},  followed by the description of the stellar models considered and the set-up for our simulations in Sec.~\ref{sec:setup_coll}. The results are presented in Sec.~\ref{sec:results} and the final discussion and conclusions in Sec.~\ref{sec:conclusions}.


\section{On the methods to simulate stellar mergers} \label{sec:methods_col}

To model stellar mergers involving an EMS and a less-massive star, it is essential to use a framework that captures both the orbital evolution of the system and the structural response of the EMS. As the less-massive star becomes engulfed within the EMS envelope, the numerical framework must resolve the inspiral dynamics, account for the work done by hydrodynamical and gravitational drag forces, and follow the resulting changes in the EMS’s internal structure and potential mass-loss. For this purpose, we employ the \texttt{StellarInspiral1D}\footnote{The code will be publicly available upon the publication of this manuscript.} code \citep[][]{fragos2019complete}, which couples the orbital evolution with stellar physics to track a star's reaction throughout the inspiral with another, lower-mass star. This code is built as an external module to \texttt{MESA} \citep{mesa1,mesa2,mesa3,mesa4,mesa5,mesa6}, extending the physics to track stellar inspirals and common envelope evolution events. Furthermore, we use a version of \texttt{MESA} that we modified as some of the relevant physical prescriptions for this project were not available in \texttt{MESA} (particularly in the release 23.05.1 that is the one used in this work) and cannot be added as an external module (or hook). The general description of the \texttt{StellarInspiral1D} code is described in Sec.~\ref{sec:inspiralcode} and the physics we incorporated in \texttt{MESA} is described in Sec.~\ref{sec:addedtomesa}.

\subsection{The \texttt{StellarInspiral1D} code} \label{sec:inspiralcode}

The \texttt{StellarInspiral1D} framework considers the interaction between two stellar objects by modeling one as a single-star \texttt{MESA} model and the other as a non-evolving spherical object. This approach was considered for cases where an extended star interacts with either a non-extended star or compact object such as a neutron star or a black hole \citep[see][as an example]{fragos2019complete}. In the case of a collision between an EMS and a less-massive star, the latter hereafter referred as the companion, we expect the EMS to have a radius three to four orders of magnitude larger than that of the companion \citep[see][]{ramirezga2025}. Thus, the previous assumption also holds for the cases regarding this work. We assume the mass of the companion is negligible with respect to the mass of the extended star, as this framework does not account for the presence of the companion star in the gravitational potential of the system \citep[see][for a possible approach on how to take into account the gravitational field of the companion]{2024A&A...683A..65B}. 

For simplicity, the companion star is modeled as a non-evolving object and it is necessary to set its mass and radius as fixed quantities, $M_{\rm comp}$ and $R_{\rm comp}$ respectively. The \texttt{StellarInspiral1D} code requires an initial configuration for the system, allowing us to freely choose the eccentricity and semi-mayor axis of the two-body system. The companion's trajectory in the $r$-$\theta$ plane is computed in the reference frame fixed at the center of mass of the EMS. The equation of motion that describes the orbit between both stellar objects is given by \citep[see][]{ginat2020gravitational}:
\begin{equation}
    \mu \ddot{\mathbf{r}} = -\frac{G\left(M_{\text{EMS}}(r) + M_{\rm comp}\right)\mu}{r^3} \hat{\mathbf{r}} - \mathbf{F}_{\text{drag}}(\mathbf{r}, \mathbf{v}) + \text{Post-Newt.},
    \label{eq:eqmotion}
\end{equation}
where $r$ is the separation between center of mass the EMS and the one of the companion ($\mathbf{r} = r\ \hat{\mathbf{r}}$), $M_{\text{EMS}}(r)$ is the enclosed mass of the EMS within separation $r$, and $\mu=M_{\text{EMS}}(r) M_{\rm comp} /(M_{\text{EMS}}(r) + M_{\rm comp})$ is the reduced mass. The term $F_{\text{drag}}$ corresponds to the drag force that is relevant when the companion star is engulfed in the EMS envelope, while the final term accounts for post-Newtonian corrections of the order 2.5.

The hydrodynamical and gravitational drag force, following \cite{ostriker1999dynamical} and \cite{binney2008}, is computed as
\begin{equation}
    \mathbf{F}_{\rm drag}(\mathbf{r}, \mathbf{v}) = - \frac{\pi\ \rho \ R_{\rm acc}^2\ \mathbf{v}^2}{\mu}\ \hat{\mathbf{v}}  \times \left\{
    \begin{array}{ll}
        \ln \left(\frac{1 + \mathscr{M}}{1 - \mathscr{M}} e^{-2 \mathscr{M}}\right), & \mathscr{M} > 1, \\
        \ln \left(\Lambda^2 - \frac{\Lambda^2}{\mathscr{M}^2}\right), & \mathscr{M} < 1,
    \end{array} \right.,
    \label{eq:fdrag}
\end{equation}
where $\mathbf{v}$ is the velocity of the companion in the considered reference frame, $\rho$ is the local EMS density at $r$, $\mathscr{M} = |\mathbf{v}|/c_{\text{\rm sound}}(r)$ is the Mach number, defined in terms of the local sound speed. $R_{\text{acc}}$ is the companion's accretion radius defined in terms of the geometric and gravitational cross-section radii as
\begin{equation}
    R_{\text{acc}} = \text{max} \left\{R_{\rm comp} \ ,\   \frac{2 G M_{\rm comp}}{v^2 + c_{\text{\rm sound}}^2} \right\}\ .
    \label{eq:r_acc}
\end{equation}
 The Coulomb logarithm $\Lambda$ is expressed as a function of the $R_{\text{acc}}$ and the convective core radius of the supermassive star \citep[see][]{binney2008,ginat2020gravitational}.

The companion’s trajectory, described by Eq.\ref{eq:eqmotion}, is solved explicitly within the \texttt{MESA} solver's time-step. The code uses a modern implementation of the Dormand-Prince 8(5,3) solver\footnote{See \href{https://github.com/jacobwilliams/dop853}{the repository of the DOP8(5,3) solver by Jacob Williams}.} based on the algorithm by \cite{hairer1993solving}. The \texttt{MESA} solver time step in our simulations is restricted to a maximum of 0.1~yr to prevent abrupt changes in the system's orbital configuration.

The work done by the hydrodynamical drag force, $\mathbf{F}_{\rm drag}\ \bullet\ \Delta \mathbf{r}$, is deposited as an additional heat source in the EMS's envelope at the companion’s position within one accretion radius, i.e. from $r-R_{\rm acc}$ to $r+R_{\rm acc}$. While the change in the orbital energy and angular momentum as a consequence of $\mathbf{F}_{\rm drag}$ is tracked, no transfer of orbital angular momentum to the supermassive star is considered in this study.

The \texttt{StellarInspiral1D} is adapted to use \texttt{MESA}'s implicit Riemann solver for Euler equations of hydrodynamics for an inviscid fluid \citep[see][]{mesa4}. The advantage of using this solver lies on the ability to resolve contact discontinuities, such as shocks, without requiring explicit artificial viscosity \citep{toro1994restoration}, contrary to the scheme described in \citet[][]{mesa3}. Artificial viscosity can damp the EMS envelope’s kinetic energy and compromise the potential mass-loss induced by the stellar collision. We use the adaptive mesh refinement (AMR) functionality that \texttt{MESA} provides for the usage of the Riemann solver \citep[][]{mesa3} instead of its default mesh refinement scheme. We adopt the same settings for \texttt{MESA}'s AMR as in \cite{farag2022resolving}, but target $~3000$ cells for the resolution of the EMS stellar model.


\subsection{Extending the \texttt{MESA} code} \label{sec:addedtomesa}

In the previous section we discussed how the \texttt{StellarInspiral1D} code extends \texttt{MESA} in order to consider the interaction between an extended envelope and a compact star. This is possible by taking advantage of the modular nature of \texttt{MESA}. For this study we identified the need for some physical prescriptions to enhance the physical accuracy of simulations like the ones considered by the \texttt{StellarInspiral1D} code. Such modifications involve the re-implementation of the flux-limited radiation transport prescription by \cite{levermore1981flux} as discussed in Appendix~\ref{sec:fld}; limiting convection velocities as discussed in Appendix~\ref{sec:limitvconv}; and extrapolation of opacity tables for low temperature and density regions, see Appendix~\ref{sec:opacities}. The modifications described here were performed on the release 23.05.1 of \texttt{MESA}\footnote{This code can be accessed in a \href{https://github.com/JRGarza/mesa/tree/r23.05.1-kap_extrapolation_lowR}{separate fork} of the \texttt{MESA} Hub repository.}.

\section{Setting up the stellar merger simulations} \label{sec:setup_coll}

In this section, we describe the final structural adjustments applied to the EMS models after their creation, in preparation for the hydrodynamic modeling of the stellar interaction with the companion. We also outline the initial and termination conditions for the simulations. 

\subsection{Building models of extremely massive stars} \label{sec:assumptions_s1}

To obtain a proxy for the internal structure of the EMSs, we follow the same process as \citet{ramirezga2025}, building such stellar models trough high mass-accretion rates by considering the ten-fold of the rate prescription for star formation by \cite{haemmerle2019stellar}, postulated in terms of the works by \cite{Churchwell1999,henning2000massive,behrend2001formation}:

\begin{equation}
\dot{M}_{\mathrm{acc}} = \frac{f}{1 - f} \cdot \dot{M}_{\mathrm{out}} \times 10,
\quad
\left\{
\begin{array}{ll}
f = 1/3 & \text{if } M \leq 5\ {\rm M}_\odot, \\
f = 1/11 & \text{otherwise}
\end{array}
\right.
\label{eq:massacc}
\end{equation}
Here, $\dot{M}_{\mathrm{out}}$ is the mass-loss rate driven by the protostar’s bolometric luminosity:
\begin{equation}
\log\left(\frac{\dot{M}_{\mathrm{out}}}{{\rm M}_\odot}\right) = -5.28 + 0.752 \log\left(\frac{L}{{\rm L}_\odot}\right) + 0.0278 \log^2\left(\frac{L}{{\rm L}_\odot}\right). 
\label{eq:acc_rate}
\end{equation}

The EMS model is assembled starting from a 0.7~\Msun pre-main-sequence star. The mass of the model increases by cold gas accretion following the Eq.~\ref{eq:massacc} until it reaches the desired mass. We consider EMSs with masses of 1000, 3000 and 5000 M$_\odot$. All our EMS models are computed for a metallicity of [Fe/H] = -2.10 corresponding to a hydrogen fraction of $X = 0.7296$ and a metal fraction of $Z = 10^{-4}$. As well, all EMS models are non rotating.

The accretion rate given by Eq.~\ref{eq:massacc} is used to reach the desired masses for the EMS before important central H-burning occurs \citep[for justification and details see][]{ramirezga2025}. We adopt this variable accretion rate over a constant one for numerical reasons rather than considering it as an accurate physical assumption for the actual formation of EMS in massive and compact star clusters. Effectively, we require a prescription that provides a sufficiently low accretion rate when the mass of our model is still small, to avoid convergence issues, while quickly ramping up to higher values as the mass of the model increases. The accreted material has the same composition as the one in the EMS surface.

\firstrev{In this work we employ the mixing length theory (MLT) of convection by \cite{mihalas1978stellar} with $\alpha_{MLT} = 1.93$ \citep{2016ApJ...823..102C,2023ApJS..264...45F}, adopting the same scheme as in the "MLT" models by \cite{ramirezga2025}. In the latter study, the possible uncertainties on the stellar structure are investigated by other sets of models where the stellar superadiabaticity is artificially reduced given the MLT++ prescriptions by \cite{mesa2,mesa6}}. The MLT++ schemes represent a numerical engineering approach for \texttt{MESA} to deal with the evolution of massive stars. Such schemes are not compatible with a hydrodynamic solution of the stellar structure. They artificially enhance the convection flux, rendering this type of simulation numerically unstable. The adjustments are not limited by any timescale linked to the hydrodynamic solver, and therefore, \firstrev{we do not produce any models with such schemes}. For a more in depth overview on the EMS models considered here, refer to \cite{ramirezga2025} "MLT" models.

\subsection{Relaxing the Stellar Structure}

Once the EMS model reaches the desired mass, following sec.~\ref{sec:assumptions_s1}, the cold gas accretion is halted. In this initial phase, the EMS structure is computed using the hydrostatic \texttt{MESA} solver. Before initiating the collision simulations, the EMS structure must be relaxed to eliminate numerical features inherited from the accretion phase that could otherwise affect the merger dynamics. Additionally, the numerical solver must be switched from the hydrostatic to the hydrodynamic Riemann solver. The relaxation process is performed in two distinct steps:
\begin{enumerate}
    \item \textbf{Post-accretion relaxation:} Accretion is halted. At this stage the Newton hydrodynamic solver, without artificial viscosity, is used instead of the hydrostatic one. For numerical reasons we first switch to the Newton hydro solver instead of the Riemann one. As now the EMS model is not assumed to be in hydrostatic equilibrium, the EMS is relaxed numerically to remove dynamical perturbations in its structure induced by the change of numerical solver. The relaxation time is not sufficient for the EMS's thermal structure to adjust or for the star to reach hydrostatic equilibrium again. Dynamical and analytical studies indicate that the typical time between collisions in dense stellar environments, such as proto globular clusters or nuclear star clusters, is $\lesssim 10^3$ yr \citep[e.g.,][]{gieles2018concurrent,ramirezga2025,rantala2024frost,rantala2025frost}; that is shorter than the EMS's timescale to reach hydrostatic equilibrium, i.e. $\sim 10^4$ yr for the models considered in this work. Therefore, thermal perturbations remain un-relaxed {between successive collisions}, and the star is expected to remain  bloated. To keep the EMS structure bloated, the model is evolved for only 100 years, approximately one hundredth of its thermal timescale. 
    
    \item \textbf{Structural relaxation:} The Riemann solver is activated to describe the stellar structure. Flux-limited radiation transport is enabled to improve the treatment of radiative losses; as well allowing to set the boundary condition by reducing the outer optical depth from $2/3$ to $2/3 \times 10^{-1}$ (see Appendix~\ref{sec:fld} for more context). Additionally, convection velocities are limited to $\leq 0.1\ c_{\rm sound}$. To relax any perturbations induced by those changes, the duration of this relaxation step is 50 years, i.e. $\gtrsim 5$ times the EMS's dynamical timescale depending on its mass.
\end{enumerate}

A major assumption on the "structural relaxation" step is related to the surface boundary condition of the EMS, where $\tau_{\rm surf} = 2/3\  \times\ 10^{-1}$ is set as the outermost optical depth for the simulated model (the default value of $\tau_{\rm surf}$ in \texttt{MESA} is 2/3). The temperature and pressure for the outer non-simulated portion of the star, i.e. $\tau<\tau_{\rm surf}$, is set by the surface boundary condition based on the Eddington gray atmosphere model \citep[see][as an example for the derivation of this model]{weiss2004cox}. The Eddington gray atmosphere model assumes that the optically thin layers of the star are in hydrostatic equilibrium, i.e. that the bulk acceleration of the gas is zero. As a result, any natural motion or expansion of the stellar surface is artificially suppressed by this assumption. To avoid such artificial damping the outer stellar layer must reach the point where the surface pressure vanishes, i.e. $\tau \rightarrow 0$. But for numerical reasons the selected value of $\tau_{\rm surf} \sim 0.067$ is the lowest optical depth tested where no convergence issues occur at the stellar surface. We raise the awareness that our results involving the motion of the EMS surface will be affected by the mentioned limitations (see \S~\ref{sec:10_70_cases}).

Throughout both relaxation phases and the merger simulations, convection across the front of supersonic shock waves is suppressed. This transport mechanism is disabled when the change in Mach number between adjacent cells exceeds $0.03$, as in \texttt{MESA}'s default settings, to ensure the shock face remains radiative.

The structural differences in the EMS before and after the relaxation steps are minimal, but performing these steps is crucial to ensure numerical stability during the subsequent merger simulations.

\subsection{Initial and Termination Conditions for the Stellar Inspiral}

After both relaxation phases, we set the initial conditions for the collisions using the \texttt{StellarInspiral1D} framework. We consider companion masses of 10, 15, 20, 30, 40, 50, 60, and 70~\Msun. The corresponding radii for these companions are obtained from gas accretion models constructed using the unmodified scheme of \cite{haemmerle2019stellar}, specifically by applying Eq.~\ref{eq:massacc} divided by 10, and adopting the same composition as the EMS models. An overview on the relevant parameters for the companion are presented in Appendix~\ref{sec:companion_params}.

Following similar assumptions to those in \cite{ramirezga2025}, all configurations begin in an initial circular orbit with a separation set to $0.8\ R_{\rm EMS}(t=0)$. This separation is chosen to avoid the companion exit the EMS during the inspiral as the outer layers pulsate due to the orbital energy injected, as this behavior was compromising the convergence of the solver.

The companion’s trajectory is followed until one of the following conditions is satisfied:
\begin{itemize}
    \item The companion fills its Roche lobe, computed using the formula by \cite{eggleton1983approximations} in terms of $M_{\rm comp}$ and $M_{\rm EMS}(r)$;
    \item The temperature of the EMS at the companion's position exceeds the companion’s virial temperature, $T_{\rm vir, comp}$.
\end{itemize}
The virial temperature is defined as the characteristic temperature at which the companion's gravitational potential energy and thermal kinetic energy are in equilibrium, establishing a relationship between its mass, radius, and internal temperature (see Appendix~\ref{sec:companion_params} as well as \cite{ramirezga2025} for more context). When either of these conditions is met, the companion is considered to dissolve within the EMS's interior. This event will hereafter be referred to as the merger event. Once a merger condition is reached, the companion star is removed from the simulation, and the orbital evolution is halted. No further energy is deposited into the EMS envelope. For numerical simplicity, the companion's mass is not added to the EMS envelope.

All simulations are tracked until they either reach 5000~yr of evolution, consume approximately 600 CPU hours, or terminate due to solver failure. The results of these simulations are discussed in the following section.

\begin{figure*}
    \centering
    \begin{subfigure}{3.01in}
        \centering
        \includegraphics[width=3.01in]{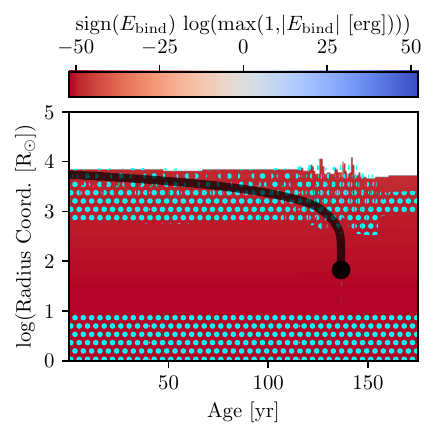}
   \includegraphics[width=3.01in]
   {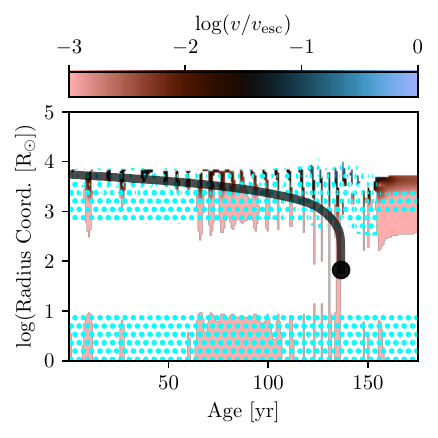}
        \caption{}
        \label{fig:insp_1e3_10a}
    \end{subfigure}
    \hspace{0.1in}
    \begin{subfigure}{3.5in}
        \centering
        \includegraphics[height=6in]{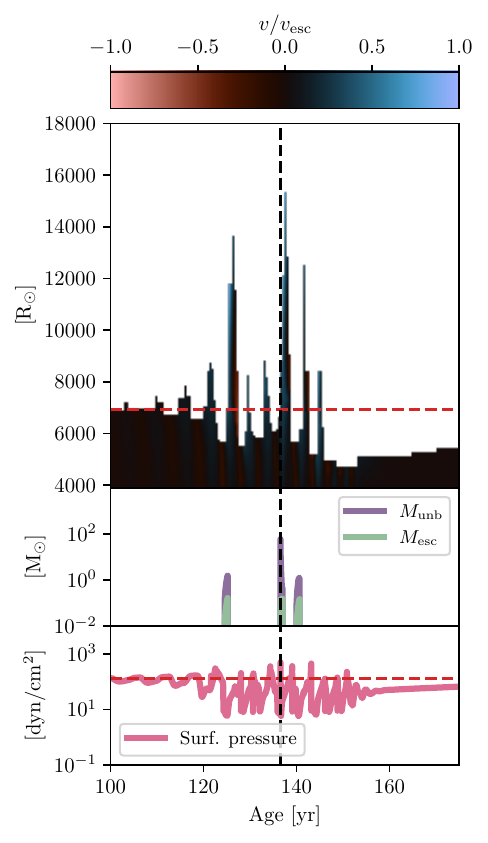}
        \caption{}
    \label{fig:insp_1e3_10b}
    \end{subfigure}
    \caption{Inspiral evolution of a 1000~\Msun\ EMS with a 10~\Msun\ companion. Left panel (Fig.~\ref{fig:insp_1e3_10a}): structural and dynamical evolution of the 1000~\Msun\ EMS during inspiral with a 10~\Msun\ companion. {Top:} EMS Kippenhahn diagram, in terms of the radius coordinate with respect to time, with the signed logarithm of the binding energy encoded by the background color; {Bottom:} same but for $\log(v/v_{\rm esc})$. In both panels  cyan dotted regions mark convection zones. The black solid line represents the position of the companion, the black dot marks the merger event. Right panel (Fig.~\ref{fig:insp_1e3_10b}): Zoom-in on the merger event. {Top:} EMS radius with $v/v_{\rm esc}$ color-map, the value of the radius at $t=0$ is shown by the red dashed line; {Middle:} unbound ($M_{\rm unb}$) and escaping ($M_{\rm esc}$) masses. {Bottom:} surface pressure, its initial value is shown by the red dashed line. In all three plots the time when the merger event is met is marked by the black vertical line.
}
    \label{fig:inspiral_evol_10}
\end{figure*}

\section{Results} \label{sec:results}

This section presents the impact of stellar mergers on the structure and mass-loss of EMSs. To establish a baseline for interpreting the results, we begin in Sec.~\ref{sec:two_case_study} with a detailed analysis of two representative cases: stellar mergers involving a 1000~\Msun\ EMS with 10 and 70~\Msun\ companions. These examples highlight the key physical processes driving mass ejection and structural changes. Sec.~\ref{sec:masloss_across_config} presents the resulting mass-loss estimates from all our simulations. Finally, in Sec.~\ref{sec:across_ecc}, given the results of the hydrodynamic simulations methodology, we provide mass-loss estimates for a broader set of initial orbital configurations, as well as a criterion to determine if a two-star interaction will lead to a merger or a scattering event, aiming to present results that can be implemented in dynamical simulations of stellar clusters.

\subsection{Stellar Mergers Between Low and High Mass Companions, a Two Case Study} \label{sec:two_case_study}

We first analyze the two cases where companions with 10 and 70~\Msun\  masses merge with a 1000~\Msun\ star. Both configurations are particularly informative toward understanding the EMS's evolution through a merger.

\subsubsection{Evolution of a 1000~\Msun\ EMS merging with a 10~\Msun\ companion} \label{sec:10_70_cases}

Let us focus on the inspiral evolution of the merger between a 1000~\Msun EMS and a 10~\Msun companion star. Fig.~\ref{fig:inspiral_evol_10} shows the system's evolution across time. The top panel of Fig.~\ref{fig:insp_1e3_10a} presents the Kippenhahn diagram showing the evolution of the EMS's binding energy (showing the value $\text{sign}(E_{\rm bind}) \times \log(\max(1,|E_{\rm bind} / [{\rm erg}]|))$) across its radial coordinate and time. The bottom panel of Fig.~\ref{fig:insp_1e3_10a} shows a similar diagram with the logarithm of the bulk velocity relative to the local escape speed, $\log(v/v_{\rm esc})$. In both plots the position of the companion is shown with the solid black line.

As can be seen in Fig.~\ref{fig:insp_1e3_10a}, most of the EMS remains gravitationally bound, as indicated by the persistent negative binding energy. However, the EMS reacts to the injected energy by pulsating with a period of approximately 6 years, close to its dynamical timescale of 5.8 years \citep[see][as a similar behavior is experienced in 1D common envelope simulations]{clayton2017episodic}. These pulsations are evident by tracking the evolution of the stellar surface across time. As the companion approaches the merger condition, the amplitude of these oscillations increases. After this point, orbital energy injection ceases, and the subsequent evolution is governed by the EMS's hydrodynamic response. The EMS expands, reaching a radius of approximately twice its initial one. 

\begin{figure*}
    \centering
    \begin{subfigure}{3.01in}
        \centering
        \begin{minipage}{\linewidth}
            \centering
            \includegraphics[width=3.01in]{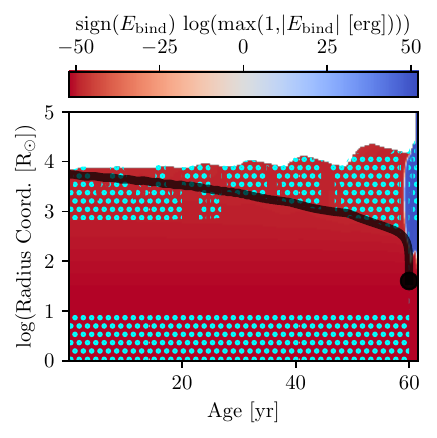}
            \includegraphics[width=3.01in]{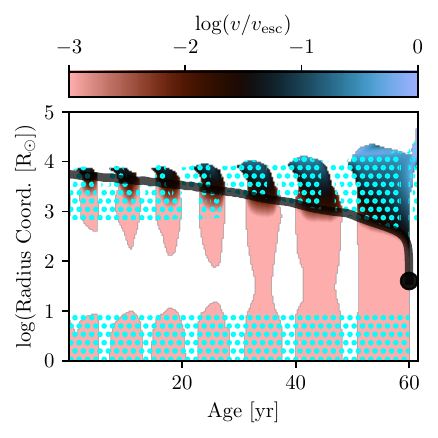}
        \end{minipage}
        \caption{}
        \label{fig:insp_1e3_70a}
    \end{subfigure}
    \hspace{0.2in}
    \begin{subfigure}{3.5in}
        \centering
        \includegraphics[height=6in]{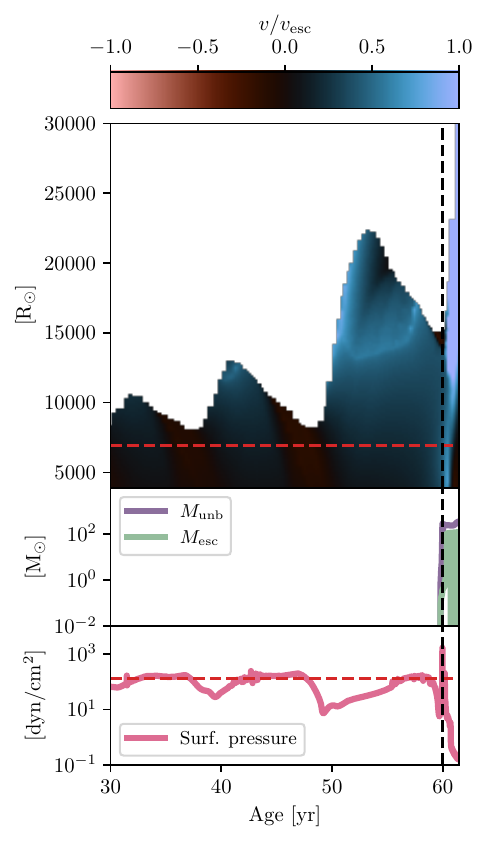}
        \caption{}
        \label{fig:insp_1e3_70b}
    \end{subfigure}

    \caption{Inspiral evolution of a 1000~\Msun\ EMS with a 70~\Msun\ companion until the simulation crashes. Left panel (Fig.~\ref{fig:insp_1e3_70a}): structural and dynamical evolution of the 1000~\Msun\ EMS during inspiral with a 70~\Msun\ companion. {Top:}EMS Kippenhahn diagram, in terms of the radius coordinate with respect to time, with the signed logarithm of the binding energy encoded by the background color; {Bottom:} same but for $\log(v/v_{\rm esc})$. In both panels  cyan dotted regions marks convection zones. The black solid line represents the position of the companion, the black dot marks the merger event. Right panel (Fig.~\ref{fig:insp_1e3_70b}): zoom-in on the merger event. {Top:} EMS radius with $v/v_{\rm esc}$ color-map, the initial radius value is shown by the red dashed line; {Middle:} unbound ($M_{\rm unb}$) and escaping ($M_{\rm esc}$) masses. {Bottom:} surface pressure, its initial value is shown by the red dashed line. In all three plots the time when the merger event is met is marked by the black vertical line.}
    \label{fig:insp_1e3_70}
\end{figure*}

In the bottom panel of Fig.~\ref{fig:insp_1e3_10a}, it is shown that after the merger condition is reached, the outer envelope accelerates outward, increasing the amplitude of the stellar pulsations. However, after a few decades, the surface motion decelerates, reducing the amplitude of the oscillations.

To investigate the mass-loss induced by the stellar merger, we define two diagnostic quantities: 
\begin{itemize}
    \item The unbound mass, $M_{\rm unb}$, corresponding to the outermost region of the EMS with positive binding energy, satisfying the condition
    \begin{equation}
        M_{\rm unb} = M_{\rm EMS} - M_{\rm low,unb}, \text{ where } \int^{M_{\rm EMS}}_{M_{\rm low,unb}}\ \varepsilon_{\rm bind}\ dm \geq\ 0.
    \end{equation}
    Here $\varepsilon_{\rm bind} = \varepsilon_{\rm grav} + \varepsilon_{\rm int} + \varepsilon_{\rm kin} $ is the specific binding energy, that is the sum of the specific gravitational, internal, and kinetic energies respectively, the injected orbital energy is already considered in the those terms. $M_{\rm low,unb} \leq M_{\rm EMS}$ is the innermost mass coordinate where the stellar surface remains unbound.
    \item The escaping mass, $M_{\rm esc}$, given by material in the outer EMS layers that satisfy
    \begin{align}
    M_{\rm esc} &= M_{\rm EMS} - M_{\rm low,esc},  \nonumber \\ 
    &\text{ where } \int^{M_{\rm EMS}}_{M_{\rm low,esc}} \varepsilon_{\rm bind}\, dm \ge 0
    \ \text{ and }\  \varepsilon_{\rm kin} \ge \varepsilon_{\rm int}.
    \end{align}
     The last condition ensures that $M_{\rm esc}$ only considers the portion of the star that is unbound by its bulk motion, i.e. whose velocity reaches or exceeds the local escape velocity. As in the definition of $M_{\rm unb}$, the injected orbital energy is already considered in the binding energy of the star.
\end{itemize}

  \firstrev{It is important to note that 
  $M_{\rm esc}$ is the mass that is already escaping the system's gravitational potential, as its kinetic energy is greater that its gravitational one. In contrast,  $M_{\rm unb}$ quantifies the material that can potentially be ejected, as the binding energy of such material can be positive due to its internal energy while its velocity has not exceeded yet the local escape velocity.} As the binding energy of the EMS will evolve due to the time dependence of the orbital injected energy, both $M_{\rm esc}$ and $M_{\rm unb}$ are time dependent. The main goal of this case study is to determine the values of $M_{\rm esc}$ and $M_{\rm unb}$.

The time evolution of the EMS radius, $M_{\rm esc}$, $M_{\rm unb}$, and EMS surface pressure are illustrated in Fig.~\ref{fig:insp_1e3_10b}. The top panel shows the EMS radius with the colormap encoding the fraction $v/v_{\rm esc}$, the middle panel tracks the evolution of $M_{\rm esc}$ and $M_{\rm unb}$, and the bottom panel presents the surface pressure evolution. As the companion approaches the merger condition, both $M_{\rm unb}$ and $M_{\rm esc}$ become significant, reaching values between $\sim1$ and $\sim60$~\Msun\ for $M_{\rm unb}$, and around $0.1$~\Msun\ for $M_{\rm esc}$. \firstrev{The maximum value reached by the unbound mass is a consequence of the binding energy of the EMS outer layers and the injected orbital energy by the companion. The EMS’s outermost $\sim70$~\Msun have a binding energy that is about two orders of magnitude lower than that of the deeper stellar layers. Moreover, the energy injected by the companion is almost an order of magnitude larger than the binding energy of this outer material. In contrast, the value of the escaping mass remains sensitive to how the EMS responds by expanding its envelope and converting internal energy into gravitational and kinetic energy.}

A comparison between the top and middle panels \firstrev{of Fig.~\ref{fig:insp_1e3_10b} allows us to understand the evolution of $M_{\rm unb}$ and $M_{\rm esc}$}. The phases during which $M_{\rm unb}>0$ and $M_{\rm esc}>0$ correspond to the periods of high-amplitude oscillations. However, there are also intervals during which both masses drop to zero and the EMS radius contracts. This indicates that some material that had initially reached or exceeded the escape velocity is decelerated and becomes bound again. This behavior is a consequence of the boundary conditions imposed by the Eddington gray atmosphere, under which the surface pressure is computed assuming a hydrostatic atmosphere. In a more realistic treatment, the surface pressure should vanish once the local escape velocity is reached. As shown in the bottom panel, however, the surface pressure decreases by only about one order of magnitude from its initial value. The EMS pulsations are damped within a few decades, not only as a consequence of the hydrostatic atmosphere but as well it is expected that radial pulsations to be damped as a consequence of the backward differences applied by the implicit numerical integration scheme used by \texttt{MESA} \citep[see ,e.g.,][for context on the damping of radial pulsations by numerical integration schemes similar to the one used in this work]{heger1997pulsations,appenzeller1970evolution}. \firstrev{In general, the EMS transforms its internal energy, that contains the thermal energy injected by the companion, to gravitational and kinetic energies as it pulsates. Then, the kinetic energy gets artificially damped, loosing a fraction of the energy that can potentially drive mass-loss.} \firstrev{In this simulation, the EMS loses a maximum of $\sim0.1$~\Msun\ per oscillation period of approximately 6 years}, where such estimate relies on the physical assumptions imposed in our models. We expect that in nature, the EMS would relax its structure and stop pulsating on a timescale comparable to its thermal timescale following the orbital energy injection.

To estimate an upper limit for the total escaping mass driven by EMS pulsations, we estimate the time it takes for the EMS to thermally relax after the stellar merger. Prior to the perturbation by the companion, the EMS luminosity does not vary significantly over a timescale comparable to the inspiral duration. As the lost orbital energy is deposited in the envelope, the difference between the instantaneous EMS luminosity, $L_{\rm EMS}(t)$, and the initial luminosity, $L_{\rm EMS}(t=0)$, represents the rate of energy loss due to the injection of orbital energy. We define the luminosity fraction $f_L$ as the average fractional increase in EMS luminosity over the initial value, integrated from the beginning of the simulation to the time the merger condition is reached:
\begin{equation}
    f_L \equiv \frac{1}{L_{\rm EMS}(t=0)\, t_{\rm merger}} \int_0^{t_{\rm merger}} \{L_{\rm EMS}(t') - L_{\rm EMS}(t' = 0)\}\, dt' \ ,
    \label{eq:lum_frac}
\end{equation}
where $t_{\rm merger}$ is the time taken to reach the merger condition from the beginning of the simulation. We integrate only up to $t_{\rm merger}$ rather than the full simulation duration because different configurations have varying termination times. Additionally, the boundary conditions of the EMS atmosphere may result in an underestimated $f_L$ after the merger.

The time it takes for the EMS to relax thermally is given by
\begin{equation}
    \tau_{\rm th,insp} = \frac{-\Delta E_{\rm orb}(t_{\rm merger})}{ f_L\ L_{\rm EMS}(t=0) }\ .
    \label{eq:tau_th_insp}
\end{equation}
The expected duration of EMS pulsation induced mass-loss is therefore of order $\tau_{\rm th,insp}  - t_{\rm merger}$, as a portion of the deposited energy is radiated away before the merger condition is reached. For the case involving the 1000~\Msun\ EMS and 10~\Msun\ companion, this time difference is approximately $420$~yr~$-~137$~yr~$=~283$~yr. With a mass-loss rate of $\sim0.1$~\Msun\ per oscillation cycle of $\sim6$~yr over 283 years, the total escaping mass is estimated to be $M_{\rm esc} \sim 5$~M$_\odot$. As our EMS structures are based on the models of \cite{ramirezga2025}, we compare this result to their mass-loss predictions where no mass-loss is expected for the configuration considered here. This hints that the assumptions in their semi-analytical work might underestimate the amount of collision induced mass-loss, as discussed in more details in Sec.~\ref{sec:masloss_across_config}.

\begin{figure*}[htbp]
   \centering
    \includegraphics[height=4in]{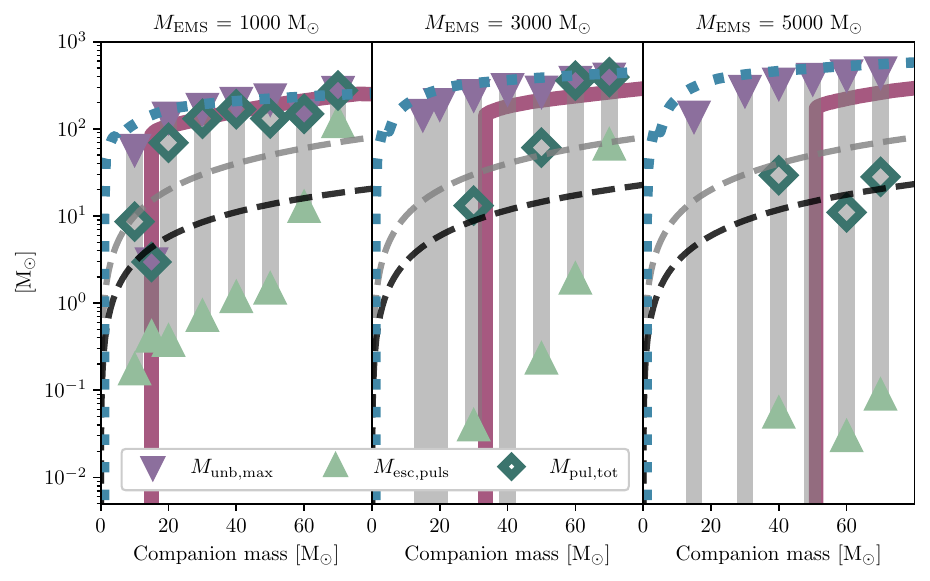} 
   \caption{Maximum unbound mass ($M_{\rm unb,max}$, purple triangles pointing down), maximum escaping mass ($M_{\rm esc,max}$, mint triangles pointing up) and predicted total pulsation mass-loss ($M_{\rm pul,tot}$, turquoise diamonds) as a function of companion mass for the three EMS models considered. The gray dashed line indicates the one-to-one relation, a predicted mass-loss below such line means that the EMS gains more mass by the stellar merger than the one it looses through it, and vice versa. The blue dotted line shows the maximum unbound mass estimation if there is no loss of injected energy during the inspiral, i.e. $\alpha=1$ at all times. The black dashed line shows the extrapolated mass-loss prescription from \citet{glebbeek2013structure}. Analytical predictions from \citet{ramirezga2025} are shown with the cherry-hue solid lines.}
   \label{fig:esc_mass}
\end{figure*}

\subsubsection{Evolution of a 1000~\Msun\ EMS merging with a 70~\Msun\ companion} \label{sec:ev_1000_70}

The resulting evolution of the stellar merger between a  1000~\Msun\ EMS and a 70~\Msun\ companion is presented as a representative case for stellar inspirals involving a high-mass companion. Fig.~\ref{fig:insp_1e3_70} shows the evolution for the corresponding system similar to Fig.~\ref{fig:inspiral_evol_10}. In the top panel of Fig.~\ref{fig:insp_1e3_70a} we observe that the binding energy of the EMS is predominantly negative but as the companion approaches the merger condition a portion of the outer envelope becomes unbound. This behavior is more noticeable for this inspiral with respect to the one involving the 10~\Msun\ companion. As the amount of deposited energy is seven times the one from the least massive companion, and the inspiral time is reduced to less than one half for the 70~\Msun\ companion, the resulting energy rate injection $\sim 15$ times larger than in the 10~\Msun companion case. 

As in the inspiral involving the 10~\Msun\ companion, the EMS  pulsates in response to the injected energy. The mean period of the oscillation for this configuration is $\sim 8$ yr. In the bottom panel of Fig.~\ref{fig:insp_1e3_70a} we present the evolution of $v/v_{\rm esc}$ fraction for the new configuration. As the companion plunges deeper in the EMS envelope, the magnitude of the bulk velocity and the amplitude of the oscillations increase. Once the merger condition is reached the EMS undergoes a rapid expansion of its outer layers increasing by nearly one order of magnitude in radius. The simulation is unable to evolve further due to numerical convergence issues, but it is likely that if further pulsations occur the amount of mass lost will increase as in the 10~\Msun merger case.

The time evolution of the EMS radius, the resulting unbound and escaping masses, as well as the surface pressure for this simulation are presented in Fig.~\ref{fig:insp_1e3_70b}. The top panel shows the EMS radius and $v/v_{\rm esc}$ increase as the companion approaches the merger event. When the merger condition is reached both $M_{\rm unb}$ and $M_{\rm esc}$ continue to increase until reaching a value of $\gtrsim10^2$~\Msun.

In the bottom panel of Fig.~\ref{fig:insp_1e3_70b}, the surface pressure is shown to decrease {by} three orders of magnitude with respect {to} its initial value as a consequence of the energy injected by the 70~\Msun companion. The low, but not null, value of the surface pressure reached as the EMS expands is one order of magnitude less than the lowest value reached by the inspiral involving the 10 M$_\odot$ companion. The motion of the outer envelope has not been decelerated completely by the time the simulation stops. Such behavior is interpreted as a consequence of the higher energy injection rate by this high mass companion with respect to the 10~\Msun one.

If we consider that the EMS will continue to eject mass similarly to the case involving the 10~\Msun companion then it is expected that a maximum of $\sim 100$~\Msun will be ejected per each 8 yr period. The inspiral thermal-timescale for this configuration is $\tau_{\rm th,insp}=592$ yr, while $t_{\rm merger} = 60$ yr. A total pulsation mass-loss can be estimated with this information, but for this configuration its value of $\sim6000$~\Msun exceeds the mass of the EMS. This is a consequence of considering that in each pulsation $100$~\Msun will be ejected as in the first one but such result must be interpreted as an upper value estimate. As a first order approximation, the total escaping mass should not exceed the maximum amount of mass that can be energetically unbound by the injected orbital energy. The theoretical estimate of such mass is obtained by considering that all deposited $-\Delta E_{\rm orb}$ is distributed just at the surface of the unperturbed EMS such that the binding energy of the external layers is exactly zero; for this configuration such mass is $\sim 300$~M$_\odot$. We will compare our mass loss estimates with calculations from other works in the following section.

\begin{figure*}[htbp]
   \centering
    \includegraphics[height=3in]{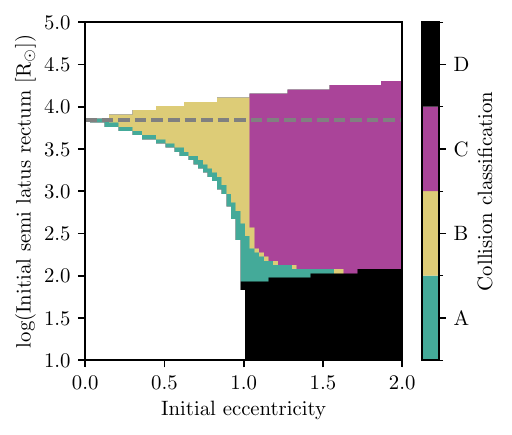} 
    \hspace{0.1in}
    \includegraphics[height=3in]{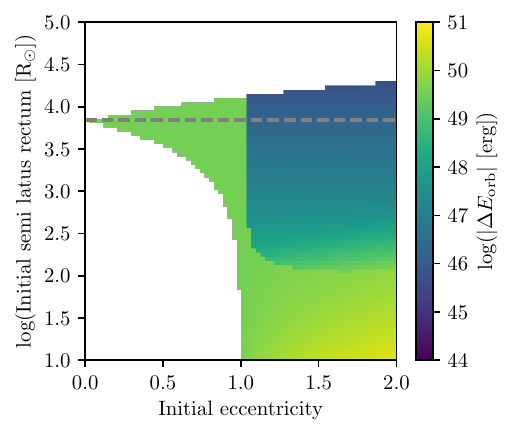} 
   \caption{Collision outcomes and deposited orbital energy for the interaction between a $1000$~\Msun\ EMS and a $70$~\Msun\ companion across initial orbital configurations. The EMS structural profile after the relaxation steps is used for trajectory integration. The initial semi latus rectum ranges from $10$ to $10^5$~R$_\odot$ and the eccentricity from $0$ to $2$. The EMS radius is indicated by the gray dashed line. {Left plot:} classification of collision outcomes (A to D, see Sec.~\ref{sec:across_ecc}) as a function of the initial semi latus rectum and eccentricity. {Right plot:} deposited orbital energy for mergers and scatterings in the absence of external interactions.}
   \label{fig:col_class}
\end{figure*}

\subsection{Resulting mass-loss across EMS and companion mass configurations} \label{sec:masloss_across_config}

The unbound and escaping masses are quantities that evolve with time and, as an impact of the assumed atmospheric boundary condition in our models and assuming the treatment of other physical processes to be accurate for our simulations, they might artificially decrease. Thus, analyzing their values at the end of the simulations will lead to an underestimation of both quantities. 
As stated in the previous section, the EMS structure reacts to the injected orbital energy by pulsating, converting the thermal energy into kinetic energy. During this process, the actual mass being lost from the EMS corresponds to the mass that escapes the gravitational potential of the star, while the energetically unbound mass that has not yet reached the escape velocity is considered as the portion of the EMS that can be lost in future pulsations. In most of our inspiral simulations, the outer envelope of the EMS exceeds its local escape velocity in the first post-merger pulsation but any subsequent mass-loss after this event is damped quickly. In order to have a global estimate for the mass-loss produced by the stellar mergers, we {introduce} two quantities:
\begin{itemize}
    \item the maximum unbound mass, across time, post-merger ($M_{\rm unb,max}$) during the inspiral evolution as a proxy for the maximum amount of mass that can be lost;
    \item the escaping mass produced by the first pulsation post-merger ($M_{\rm esc,puls}$), as a proxy of the maximum amount of mass that can be lost from each pulsation of the EMS produced by the injected energy.
\end{itemize}
In addition, we define the predicted total mass-loss produced by the EMS pulsations post-merger as
\begin{equation}
    M_{\rm puls,tot} = \min{ \left\{ M_{\rm unb,max}\ ,\  \frac{M_{\rm esc,puls}}{\tau_{\rm puls}}\ \left( \tau_{\rm th,insp} - t_{\rm merger} \right)  \right\} },
\end{equation}
where $\tau_{\rm puls}$ is the dominant period of the EMS radius oscillations before the merger event,  calculated in terms of the inverse frequency-spectrum from the fast Fourier transform of the EMS radius evolution in such time domain. As mentioned before, the value of $\tau_{\rm pul}$ is of the order of the EMS dynamical timescale. The previous definition assumes that if the motion of the outer envelope is not damped, then the EMS would produce an escaping mass of $M_{\rm esc,puls}$ per oscillation period $\tau_{\rm pul}$. This would last until the EMS manages to radiate away the total injected energy after the merger condition is reached. In addition, we consider $M_{\rm unb,max}$ as the maximum mass that can be lost by the EMS, and thus limit $M_{\rm pul,tot}$ to be $M_{\rm pul,tot}>M_{\rm unb,max}$.

\begin{figure*}[htbp]
   \centering
    \includegraphics[height=3in]{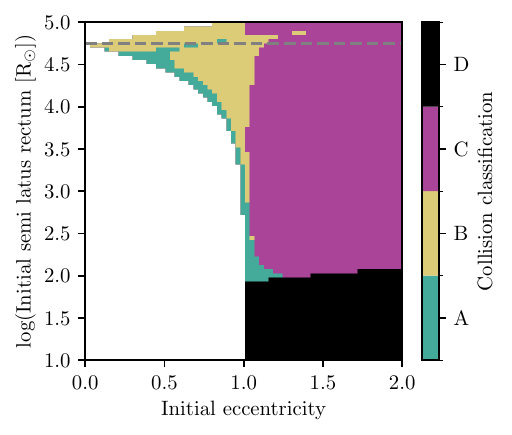} 
    \hspace{0.1in}
    \includegraphics[height=3in]{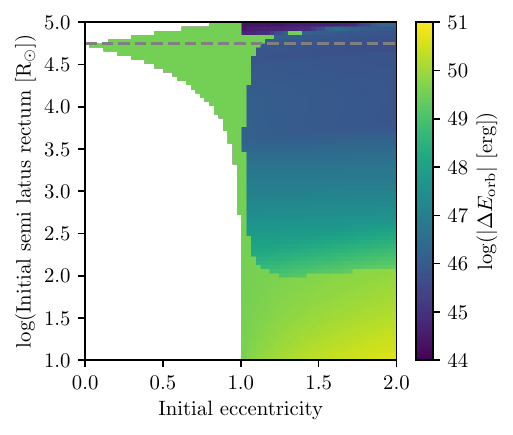} 
   \caption{Same as Fig.~\ref{fig:col_class}, but using the EMS profile at the end of the hydrodynamic collision simulation for the interaction between a $1000$~\Msun\ EMS and a $70$~\Msun\ to determine the possible outcomes of a second collision. The EMS has expanded to a radius of $\sim 5\times 10^4$~R$_\odot$, shown by the gray dashed line in both panels. {Left plot:} collision outcome classification, {right plot:} deposited orbital energy for the same configurations.}
   \label{fig:col_class_postcoll}
\end{figure*}

The values of $M_{\rm unb,max}$, $M_{\rm esc,puls}$, and $M_{\rm puls,tot}$ are shown in Fig.~\ref{fig:esc_mass} as a function of companion mass for all the EMS and companion-mass configurations considered in this work. The discussion of this plot will continue in the following sections.

\subsubsection{{Comparison with analytical mass-loss estimates}}

 To compare our results with the predictions by \cite{ramirezga2025} we need to extract the fraction of deposited orbital energy that is used to drive mass-loss, defined here as the parameter $\alpha$. The value of $\alpha$ is determined by the amount of $-\Delta E_{\rm orb}$ that has not been lost yet, e.g. has not been radiated away, by the EMS at $t_{\rm merger}$, i.e.
\begin{equation}
    \alpha = \frac{\text{total injected energy} - \text{injected energy lost}}{\text{total injected energy}}.
\end{equation}
 In the framework by \cite{ramirezga2025} the value of $\alpha$ is not calculated explicitly but is determined by the following assumptions:
\begin{itemize}
    \item the orbital energy deposited in regions where the local thermal timescale $\tau_{\rm thermal,local}(r) = E_{\rm int}(r)/L_{\rm surf}$ is shorter than the local inspiral timescale $\tau_{\rm insp} = r/\dot{r}$ is promptly radiated away and excluded from the total orbital energy injected $\Delta E_{\rm orb}$;
    \item only the orbital energy deposited above the innermost coordinate $r$ where $\alpha\ \vert E_{\rm orb}(r) \vert >  \vert E_{\rm bind}(r) \vert $ is used to unbound the mass above $r$.
\end{itemize}
In such framework, it is assumed that all the predicted unbound mass will escape. 

In our simulations, the value of $\alpha$ until merger can be estimated in terms of the amount of $-\Delta E_{\rm orb}$ and by assuming the injected energy can only be lost by being radiated away as:
\begin{equation}
    \alpha = \frac{-\Delta E_{\rm orb}(t_{\rm merger}) - f_L\ L_{\rm EMS}(t=0)\ t_{\rm merger}}{-\Delta E_{\rm orb}(t_{\rm merger})} \ .
    \label{eq:beta_def}
\end{equation}
An overview on the comparison on the values of $\alpha$ from the framework by \cite{ramirezga2025} and our results is presented in  Appendix~\ref{sec:beta_values}. 

The mass-loss estimates from the cited analytical framework are shown in Fig.~\ref{fig:esc_mass} by the cherry-hued solid line. It is observed that our simulations predict higher unbound mass compare t the analytical estimates. This is because \cite{ramirezga2025} assume that no mass can be unbound if the injected orbital energy is lower than the local EMS binding energy,   i.e. $\alpha = 0$. Consequently, no mass-loss is expected from mergers involving low-mass companions (with the minimum companion mass to induce mass loss increasing with the mass of the EMS). From our simulations, however, a negligible value of $\alpha$ is produced by systems where $t_{\rm merger} \gg \tau_{\rm th,insp}$ as expected from Eq.~\ref{eq:beta_def}, and if one substitutes the definition of $\tau_{\rm th,insp}$ (Eq.~\ref{eq:tau_th_insp}) it is obtained that
\begin{equation}
    \alpha = 1 - \frac{t_{\rm merger}}{\tau_{\rm  th,insp}}\ .
    \label{eq:beta_def_timescale}
\end{equation}

An analytical estimate for the maximum amount of mass that can be energetically unbound is shown in Fig.~\ref{fig:esc_mass} with the blue dotted line. This estimate is determined by considering the initial EMS profile, following the same arguments as in \cite{ramirezga2025}, but assuming that $\alpha=1$, i.e. none of the orbital injected energy is lost during the inspiral and all of it is distributed to set the EMS binding energy of the surface to zero. The resulting $M_{\rm unb,max}$ from our simulations is overall consistent with this analytical estimate for high-mass companions and significant discrepancies arise for low-mass ones. Such differences are, mainly, a consequence of not taking into account orbital energy losses during the inspiral.

\begin{figure*}[htbp]
   \centering
    \includegraphics[height=2.9in]{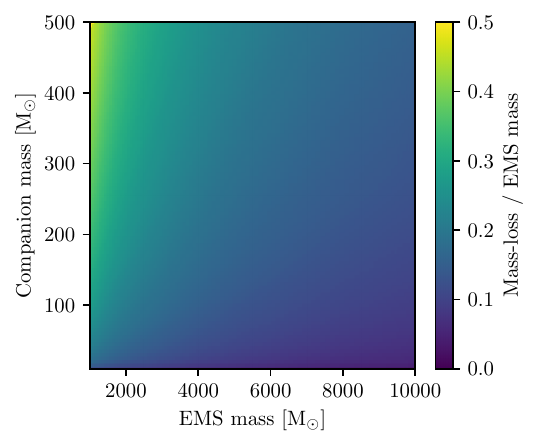} 
    \hspace{0.0in}
    \includegraphics[height=2.9in]{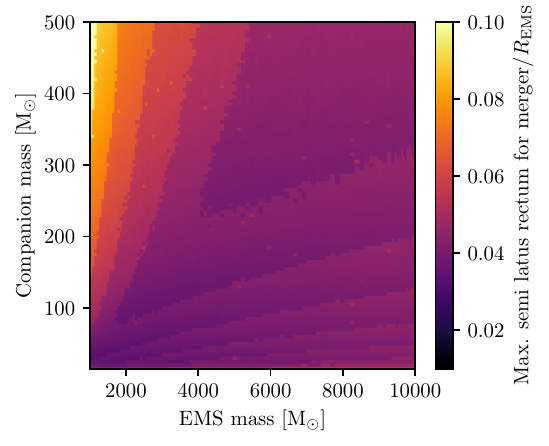} 
   \caption{Relative estimates of merger-induced mass-loss and the semi latus rectum threshold between stellar mergers and scatterings, for orbits with initial eccentricity of 1.01. Considering EMS masses from $10^3$ to $10^4$~\Msun\ and companion masses from 5 to 500~\Msun. {Left plot:} ratio of the maximum merger-induced mass-loss to the EMS mass. {Right plot:} ratio of the maximum initial semi latus rectum for a stellar merger to the EMS radius.}
   \label{fig:max_massloss_lr}
\end{figure*}

It is relevant to point out that the resulting values of $\alpha$ from our simulations are  dependent on the choice of initial conditions. All of our numerical experiments start from a circular orbit. It is expected that for eccentric configurations the ratio ${t_{\rm merger}}/{\tau_{\rm  th,insp}}$ will change for the same EMS and companion masses, resulting in different values for $\alpha$.  If $\alpha$ gets closer to unity, the maximum unbound mass estimate will become a better representation of the possible mass-loss from such stellar mergers.

Regarding the total mass-loss estimation from the pulsation behavior, for the 1000~\Msun EMS the resulting $M_{\rm esc,puls}$ is consistent with the maximum unbound mass. It is also observed that the match between $M_{\rm esc,puls}$ and $M_{\rm unb,max}$ shows a growing discrepancy for the configurations with more massive EMS models. We interpret {this} to be a {consequence of} the assumed hydrostatic atmosphere condition, as for more massive EMS models will require a higher injected energy to expand the star, and such expansion will be damped by the hydrostatic atmosphere leading to an underestimated $M_{\rm esc,puls}$ and therefore an underestimated $M_{\rm pul,tot}$.

\subsubsection{{Comparison with mass-loss estimates from other hydrodynamic simulations} }

We compare our results to the mass-loss prescription from \citet{glebbeek2013structure} based on their hydrodynamic simulations for head-on collisions of lower-mass stars, i.e. from 5 to 40 M$_\odot$. As the dynamical configuration, as well as the physical solution of the stellar merger, is different for that work with respect to ours we perform a comparison of both to raise awareness on the impact of the considered physical assumptions on the resulting mass-loss estimates from both studies.

The predicted mass-loss by \citet{glebbeek2013structure} is shown in Fig.~\ref{fig:esc_mass} as the black dashed line. Such results predict a maximum mass-loss of $10\%$ of the total mass of the system, but such high mass-loss is experienced only when the value of the mass of both stars in the collision are similar, i.e. $m_{\rm star1} \simeq m_{\rm star2}$. The configurations considered in our study consist of extremely low mass-ratios, $M_{\rm comp} / M_{\rm EMS} \lesssim 0.1$, where \citet{glebbeek2013structure} predicts a mass-loss of $< 2\%$ of the system mass $M_{\rm EMS} + M_{\rm comp}$. From our results, the value of $M_{\rm unb,max}$ represent $\lesssim 20\%$ of the system mass, similarly for $M_{\rm pul,tot}$ in the case of the 1000~\Msun EMS. We attribute the high percentage of unbound mass from our simulations, and assumed mass-loss, to the structure of the EMS. The EMS envelope is mostly radiation pressure dominated as the EMS luminosity approaches the Eddington limit; as well the EMS is marginally bound as the value of the adiabatic exponent is close to $4/3$ across the star. There is no further details on the structure of the stars considered by \citet{glebbeek2013structure} but differences on the resulting mass-loss from both studies may arise from the marginally stable EMS models considered here.

\subsection{Possible merger outcomes across eccentricities} \label{sec:across_ecc}

Within the limitations of the presented numerical experiments, the final mass-loss from a stellar merger depends primarily on the total orbital energy injected, the EMS binding energy distribution, the inspiral duration, and the timescale over which the EMS can radiate the injected energy.

The \texttt{StellarInspiral1D} framework allows simulations with initially eccentric  orbits. However, tests with such configurations show that the solver timestep is strongly modulated by the net injected energy. In eccentric orbits, the loss of orbital energy increases near periastron. This reduces the timestep of the calculations by several orders of magnitude earlier in the inspiral evolution, with respect to the initial circular orbit configuration. Following these mergers would require 10--100 times more computational time than the models considered in this work, i.e. 6000 to 60000 cpu hours.

To estimate the possible outcomes of stellar interactions across orbital configurations, we adopt an alternative method where Eq.~\ref{eq:eqmotion} is solved assuming a static EMS structure during the companion’s trajectory. This allows us to determine whether a given orbital configuration leads to a stellar merger,  and to estimate the orbital energy lost during the interaction that can be used to drive mass-loss.

We characterize the initial configuration using eccentricity ($e$) and semi latus rectum ($\ell$), the latter being always positive unlike the semi-major axis ($a$). The semi-mayor axis is related to the eccentricity and semi latus rectum as $a = \ell/(1-e^2)$. Only conics intersecting the EMS radius are considered for the initial conditions. The companion’s trajectory is integrated until completing one orbit or exiting the EMS. Outcomes are classified by the final orbital parameters as:
\begin{enumerate}[label=\Alph*.]
  \item The companion never exits the EMS.
  \item The companion exits the EMS in the first orbit with $e_{\rm final}<1$. 
  \item The companion exits the EMS in the first orbit with $e_{\rm final}>1$.
  \item The merger condition is reached during the first orbit.
\end{enumerate}

For outcomes A, B, and D, the post-interaction orbit is bound and a merger will occur in the absence of external perturbations. In case B a merger is expected, although in a dense stellar cluster the companion could, in principle, interact with other stars after leaving the EMS envelope and prevent the subsequent merger. The total injected energy for these cases is:
\begin{equation}
    -\Delta E_{\rm orb,merger} = \frac{G\,M_{\rm EMS}(r_{\rm merger})\,M_{\rm comp}}{2\,r_{\rm merger}} - \frac{G\,M_{\rm EMS}\,M_{\rm comp}}{2\,a_{\rm initial}}\ .
    \label{eq:eorb_merger}
\end{equation}
Here $r_{\rm merger}$ is the separation at which the merger condition is met, $M_{\rm EMS}(r_{\rm merger})$ is the enclosed mass at that radius, and $a_{\rm initial}$ is the initial semi-major axis. 

For outcome C, the orbit remains unbound and no merger occurs; this is classified as an orbital scattering. The injected energy is then:
\begin{equation}
    -\Delta E_{\rm orb,scattering} = \frac{G\,M_{\rm EMS}\,M_{\rm comp}}{2\,a_{\rm final}} - \frac{G\,M_{\rm EMS}\,M_{\rm comp}}{2\,a_{\rm initial}}\ .
    \label{eq:eorb_scattering}
\end{equation}

To illustrate the previous methodology, Fig.~\ref{fig:col_class} shows, for a $1000$~\Msun\ EMS and a $70$~\Msun\ companion, the collision outcomes and deposited orbital energy across initial orbital configurations. The EMS profile after relaxation is used for trajectory integration. The semi latus rectum ranges from 10 to $10^5$~R$_\odot$, eccentricity from 0 to 2, and the EMS radius is indicated by the gray dashed line. Initially bound orbits ($e<1$) remain bound, as drag reduces orbital energy and angular momentum, decreasing semi latus rectum and eccentricity. For initially unbound orbits ($e>1$), mergers require a semi latus rectum $\sim 100$~R$_\odot$ ($\sim 1/100$ of the EMS radius). For $e\approx 1$, all allowed initial configurations yield bound orbits, but only A and D guarantee a merger; B requires  no external interaction to remain bound.

It is seen from the right panel of Fig.~\ref{fig:col_class} that for the mergers without external interactions (cases A, B, D) the deposited energy is nearly independent of the initial semi-major axis. Thus, mass-loss should mainly depend on the rate of orbital energy injection and the EMS response timescale. In C, the injected energy decreases steeply with increasing semi latus rectum; without a merger, mass-loss is still possible but at much lower levels given the amount of injected energy. If in a different scenario we consider case B as scattering events as only cases A and D ensure a direct merger, then the injected energy for case B will be the same as in case C.

Even though a static profile method expands the accessible orbital parameter space, it ignores the EMS's reaction to energy injection by successive collisions. As described in Sec.~\ref{sec:two_case_study}, the EMS structure will be affected depending on the companion star's mass, possibly leading to a more extended and less bound EMS. To illustrate the possible collision outcomes for a successive stellar merger of a $1000$~\Msun\ EMS with two companions of the same mass, Fig.~\ref{fig:col_class_postcoll} shows the collision outcomes for the configuration of a $1000$~\Msun\ EMS and $70$~\Msun\ companion but using the EMS profile at the end of the hydrodynamic simulation from the merger with the same mass configuration (see Sec.~\ref{sec:ev_1000_70} for the evolution EMS structure during a first merger with also a $70$~\Msun star). The EMS radius has expanded to $\sim 5\times 10^4$~R$_\odot$ (gray dashed line). With this larger structure, initially bound orbits with higher semi latus rectum can be enclosed. For initially unbound orbits, the outcome changes little: direct mergers still require $\ell\lesssim 100$~R$_\odot$. Thus, using the initial EMS profile underestimates mergers for extended bound orbits but yields similar results for unbound orbits.

With the previous methodology we can estimate the orbital energy lost during a two star collision. One can also estimate the maximum amount of mass-loss by assuming all the lost orbital energy is used to unbiund mass in the EMS outer layers. Such prediction corresponds to value of outer mass where the orbital energy injected is equal to the EMS binding energy integrated from the surface inwards. The mass-loss estimate for mergers depends on the mass of both stars as well on the initial eccentricity and semi latus rectum, while for scattering events the mass-loss will depend on the final orbital parameters as well (see Eq.~\ref{eq:eorb_scattering}).

To illustrate the predicted mass-loss for direct mergers across stellar masses we consider a fixed initial eccentricity of $e_{\rm initial}=1.01$ to reduce the dimension of the parameter space required to determine such estimate. The selection of initial eccentricity is representative of nearly parabolic encounters in stellar cluster simulations \citep[e.g. see][where $\gtrsim 90\%$ of stellar merges, given by the sticky sphere approximation, involve eccentricities from $0.9$ to $1.1$]{rantala2024frost,rantala2025frost}. As we determine the mass-loss only for direct stellar mergers, i.e. case A and D, this implies that the initial semi latus rectum is small enough to lead to such collision outcome. We determine as well such threshold value for the semi latus rectum to produce a direct merger with $e_{\rm initial}=1.01$. 

Fig.~\ref{fig:max_massloss_lr} shows the estimates of the mass-loss and semi latus rectum threshold for mergers, in units of EMS mass and radius, respectively, as a function of EMS masses ($10^3$–$10^4$~\Msun) and companion masses (5–500~\Msun). In the left panel, the mass-loss fraction reaches $\sim 50\%$ for high mass ratios ($M_{\rm comp}/M_{\rm EMS} \rightarrow 1$), and is negligible for small mass ratios ($M_{\rm comp}/M_{\rm EMS} \rightarrow 0$). In the right panel, the maximum $\ell_{\rm initial}$ for a merger with $e_{\rm initial}=1.01$ reaches at most $\sim 10\%$ of $R_{\rm EMS}$ for high mass ratios; in other cases, it depends on the orbital energy loss needed to trap the companion and on its gravitational cross section, both scaling with the masses involved. From the discussion about Fig.~\ref{fig:col_class_postcoll}, the EMS response can modify the semi latus rectum criterion, but for a high eccentricity such as $1.01$ such variations are assumed to be minimal. For the implementation of the resulting mass loss from this work in dynamical simulations with multiple stellar interactions, we recommend using the semi latus rectum criterion rather than the EMS radius to determine stellar mergers instead of applying the sticky-sphere approximation for collisions in terms of the EMS radius. For absolute values on the mass-loss and semi latus rectum threshold see Appendix~\ref{sec:absolute_massloss}.

\section{Conclusions} \label{sec:conclusions}
Aiming to understand one of the processes involved in the formation of EMSs and SMSs, this work investigates the role of mass-loss induced by stellar mergers. To accomplish that task, the evolution of circular stellar inspirals between massive and extremely massive stars is solved by the \texttt{StellarInspiral1D} framework adapted to the \texttt{MESA} stellar evolution code. The simulations performed involve 1000, 3000, and 5000~\Msun EMSs and inspiraling companions of $<100\ {\rm M}_\odot$. We find that the thermal energy deposited by the stellar companions traveling inside the EMS envelope, resulting from the hydrodynamic and gravitational drag force, perturbs the equilibrium structure of the EMS and leads to pulsations. Such pulsations increase in amplitude with the energy that has been injected, and the stellar surface can be accelerated until it exceeds the local escape velocity, at which point mass-loss is produced. The ejected mass by the pulsations becomes relevant as the companion star gets closer to the point where it merges with the EMS structure, as there the amount of orbital energy lost is maximal. 

When the companion stars merge with the EMS structure (they reach the merger condition), the deposition of the lost orbital energy ceases. After that point, the stellar pulsations are usually damped over a few dynamical timescales, i.e., on the order of $\sim10$ yr.

To estimate the total amount of mass that could be lost through the pulsations induced by the inspiral, we extract the amount of mass lost in the first expansion post-merger. Such mass-loss on the order of $\gtrsim 10^{-2}$ to $10^2\ {\rm M}_\odot$. The oscillation period is obtained by performing a fast Fourier transform of the evolution of the EMS radius over time; such value does not vary significantly during the evolution of the simulations. The pulsation period does vary across configurations; nevertheless, its value is always of the same order of magnitude as the EMS dynamical timescale. 

With the resulting pulsation mass-loss and period we conclude that inspirals involving low-mass companions produce a mass-loss rate from $\sim 10^{-3}$ to $\lesssim 10^{-1}\ {\rm M}_\odot/{\rm yr}$; the mergers involving high-mass companions produce a mass-loss rate from $\gtrsim 10^{-1}$ to $\sim 10\ {\rm M}_\odot/{\rm yr}$. 

To estimate the duration of the EMS pulsations post-merger, we estimate the timescale for the star to radiate away the total injected orbital energy as in Eq.~\ref{eq:tau_th_insp}. We consider that the deposited orbital energy is  radiated away at a rate of average increment of the EMS luminosity with respect to its initial value.
 EMS luminosity increases, on average, from $\gtrsim1\%$ for inspirals with lower-mass companions (i.e. $\sim 10$~M$_\odot$), to $\lesssim 200\%$ for high-mass ones. 

By scaling the resulting mass-loss rate from the pulsations across the time it takes for the EMS to radiate away the injected energy, after the merger, we estimate the total mass-loss from the stellar inspirals. The scaled mass-loss is consistent with the maximum mass that is energetically unbound across the inspiral evolution. The match between the two values, total ejected and unbound mass, is better for the least massive EMS. For such an EMS mass, the resulting damping for the pulsating motion, related to the physical assumptions and the numerical integration scheme, is less. It is concluded that most of the total injected energy is used to drive the ejection of mass (see Sec.~\ref{sec:masloss_across_config}). The last point is explained by the fact that the EMS models considered have an average adiabatic exponent $\simeq 4/3$, meaning that the stars are almost dynamically unstable and the perturbation by the stellar inspiral can efficiently lead to mass ejection. 

In terms of the system's mass ($M_{\rm EMS} + M_{\rm comp}$), the total pulsation mass-loss varies from $\sim1\%$ to $\lesssim 30\%$ across configurations. Such values are consistent with the predictions by \cite{ramirezga2025}, with the exception of the cases where such a study predicts no mass-loss but the hydrodynamical simulations here produce it. 

Given the computational time constraints imposed by the \texttt{StellarInspiral1D} framework, we do not perform simulations for eccentric mergers in this study. Nevertheless, we provide a semi-analytical prescription for such cases. By solving the companion's equation of motion, Eq.~\ref{eq:eqmotion}, on a static EMS profile, we discriminate between the orbital parameters that lead to stellar mergers and orbital scatterings. Given the initial and final semi-major axis for each interaction, we compute the total orbital energy lost through the work done by the drag force. By assuming all the deposited energy is used to drive mass-loss, we estimate the latter quantity. With this methodology, we find that the amount of orbital energy deposited by stellar mergers does not vary significantly across orbital configurations, implying a similar resulting mass-loss for those systems. However, the lost orbital energy is more sensitive to the initial orbital configuration when the companion is scattered by the EMS. Finally, initially bound orbits remain bound after the interaction, whereas initially unbound orbits require a semi latus rectum $\lesssim 0.1\ R_{\rm EMS,initial}$ across eccentricities to produce a merger instead of scattering the companion's trajectory (further details are discussed in Sec.~\ref{sec:across_ecc}).

Our results illustrate the importance of collision-induced mass loss as a fundamental process regulating the formation and growth of EMSs. In the future, an open task is to assess how the cumulative effect of this process influences the formation and long-term growth of EMSs and SMSs. In particular, its role in setting the relevance of EMSs across cosmic time, whether as progenitors of intermediate-mass black holes or linked to other astrophysical scenarios, remains to be addressed in future studies.

\section*{Data availability}
The data generated for this study are publicly available via Zenodo and can be accessed in the associated repository: \href{https://zenodo.org/records/18075985}{https://zenodo.org/records/18075985}.

\begin{acknowledgements}
      This work was supported by the Swiss National Science Foundation (PI Fragos, project number CRSII5\_213497). EF acknowledges support by the Yale Center for Astronomy and Astrophysics Prize Fellowship. \firstrev{JRG acknowledges N. Lahén and A. Rantala for their insightful comments on this manuscript, as well as the \texttt{MESA} community and developers for their for the valuable input to aid the progress of this work.}
\end{acknowledgements}

%
%

\bibliographystyle{aa} 
\bibliography{references.bib}

\begin{appendix} 
\input{appendix/extending_mesa}

\input{appendix/companion_star_info}

\input{appendix/beta_values}

\input{appendix/mass_loss_absolute}
\end{appendix}

%

\end{document}

%% file: appendix/extending_mesa.tex
\section{ Flux-limited radiation transport and surface boundary condition} \label{sec:fld} 

Having a consistent model for radiative transport valid in the extreme cases of optically thin and thick regimes is necessary to model the extended envelope during the stellar mergers. Then for this work, the flux-limited diffusion (FLD) theory developed by \cite{levermore1981flux} has been re-implemented in \texttt{MESA}\footnote{See \href{https://github.com/MESAHub/mesa/pull/771}{pull request 771 in the \texttt{MESA} Hub repository}.}. This feature was previously available in the code but had been removed in more recent versions. The FLD theory provides a closure relation for the moment equations of the radiation intensity.  As well, it provides a description for radiation transport on the transition between the diffusive regime (optically thick, $\tau \gg 1$) and the free-streaming regime (optically thin, $\tau \ll 1$). The zeroth and first order moments of the radiation intensity, the radiative energy density ($E_{\rm rad}$) and the radiative flux ($F_{\rm rad}$) respectively, are related as

\begin{equation}
    F_{\rm rad} = \frac{1}{3} \frac{c}{\kappa \rho} \lambda \nabla E_{\rm rad}\ .
\end{equation}
Where $\lambda$ is the flux-limiter derived by \cite{levermore1981flux}, that parametrizes the angular dependency of the radiation beam's intensity and is a function of the flux-ratio $R_{\rm fl} \equiv |\nabla E_{\rm rad}| / \kappa \rho E_{\rm rad}$. If $\tau \ll 1$ then $R_{\rm fl}\rightarrow \infty $ and the flux-limiter should asymptotically approach to zero, contrary to the case of $\tau \gg 1$ where $R_{\rm fl} \rightarrow 0$ and $\lambda \rightarrow 1$.

Unfortunately, \texttt{MESA} does not solve the radiation transport moment equations to obtain $E_{\rm rad}$ and implement FLD theory in a self consistent manner. This implementation of FLD requires a closure assumption on the value of $E_{\rm rad}$ in order to compute $R_{\rm fl}$ and then $\lambda$. The assumption is to consider $E_{\rm rad} = aT^4$ everywhere in the star  \cite[similarly to][]{bersten2011hydrodynamical}. But such assumption is only valid in the optically thick regime, leading to an underestimation of $R_{\rm fl}$ where $\tau \ll 1$ and therefore an overestimation of $\lambda$ in such regime. Even if the implemented FLD prescription is still too diffusive at $\tau<1$, it is the best option to describe radiative transport there within \texttt{MESA}'s limitations.

 \begin{figure}[!htbp]
   \centering
   \includegraphics[width=3.5in]{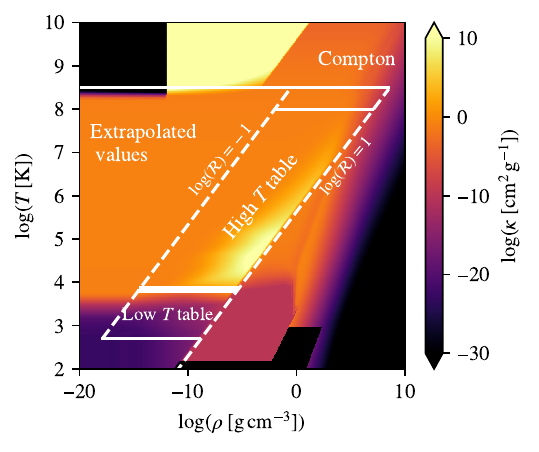} 
      \caption{Values for the stellar opacities in the $\log(T)$ vs. $\log(\rho)$ plane for $Z = 10^{-4}$. The colormap denotes the value of the logarithm of the opacity in [$\mathrm{cm}^2\,\mathrm{g}^{-1}$]. The limits for the domains of each opacity table/prescription are shown in white. The solid lines represent the boundaries in temperature, while the dashed lines represent the boundaries in $\mathcal{R}$ values. The intermediate regions between the low $T$ and high $T$ tables, as wells as between the high $T$ and the Compton opacities, are the blending zones between such domains. The low and high $T$ tables are being extrapolated when $\mathcal{R} < 10^{-1}$ or $T < 500 $K by considering the nearest-neighbor value in terms of the temperature and $\mathcal{R}$.
              }
         \label{fig:kap_regions}
   \end{figure}

\section{Limiting convective velocities} \label{sec:limitvconv}

With the motivation of avoiding supersonic convection transport, the capability to limit convection velocities has been re-implemented in \texttt{MESA}\footnote{See \href{https://github.com/MESAHub/mesa/pull/781}{pull request 781 in the \texttt{MESA} Hub repository}.}. For this work, we choose the mixing length theory (MLT) of convection by \cite{mihalas1978stellar} and \cite{kurucz1979model} with $\alpha_{MLT} = 1.93$ \citep{2016ApJ...823..102C,2023ApJS..264...45F}. Convective velocities are limited in terms of the local sound speed ($c_{\rm sound}$) to be not grater than $0.1 c_{\rm sound}$. The considered MLT prescription is only valid for non-supersonic convection velocities where the turbulent pressure is negligible with respect to the gas pressure \citep[see][for more context on subsonic convection]{weiss2004cox}.

\section{Extrapolation of the opacity tables} \label{sec:opacities}

In order to describe the interaction between radiation and matter, \texttt{MESA}'s opacity module (\texttt{kap}) contain several options for gray opacity tables that are blended in the $\mathcal{R}$-$T$ plane, where $T$ is the local stellar temperature and $\mathcal{R} \equiv (\rho/ \text{g cm}^{-3} )/ (T/10^6 \ \text{K})^3$. In a simple manner, there are {four} sets of the opacity data that are blended:
\begin{itemize}
    \item the low temperature tables (usually to describe  $\kappa$ from the temperature of formation of molecules and dust grains, ($T \sim 500$ K) until a $T$ value below the hydrogen recombination peak, $T \sim 6000$ K),
    \item the high temperature tables (for  $\kappa$  from the hydrogen recombination $T$ to $T \sim 10^8\ \text{K}$), 
    \item the Compton scattering prescription by \cite{poutanen2017rosseland} (to compute $\kappa$ for $T > 10^8\ \text{K}$ ), and
    \item the thermal conductive opacity based on the study by \cite{cassisi2007updated}.
\end{itemize}

During the development of this work we noticed that \texttt{MESA} was blending the Compton scattering opacity with the lower end of the low temperature tables, leading to an underestimation of $\kappa$ of almost 30 orders of magnitude for regions with $T < 3000 \ \text{K}$ or $\mathcal{R} < 10^{-7.5}$ \footnote{For more context in this topic, see \href{https://github.com/MESAHub/mesa/issues/814}{issue 814 in the \texttt{MESA} Hub repository}.}. Given the previous nonphysical behavior, we fixed the blending of the Compton opacity to strictly be used only if $T>10^8\ \text{K}$ or $\mathcal{R} > 10$, that correspond the $T$ and $\mathcal{R}$ limits of the tables used respectively. \\

For this study we consider the low temperature tables by \cite{ferguson2005low} and the high temperature tables by \cite{iglesias1993radiative,iglesias1996updated}. Both tables are only defined for $0.1 < \mathcal{R} < 10$ and $T > 500$ K. To avoid not having a defined opacity value for the low-temperature and low-density EMS surface as it expands, we extrapolate only the low and high temperature tables by considering the nearest-neighbor value, in the $\mathcal{R}$-$T$ plane, for the regions where $\mathcal{R} < 0.1$ or $T < 500 $K.  

The resulting opacity table used for the stellar models in this work is presented in Figure~\ref{fig:kap_regions}. The default \texttt{MESA} opacity tables are extrapolated below their minimum values for temperature and $\mathcal{R}$. The EMS models in this work only cross such lower boundary values of the default tables once the stellar surface expands due to the energy injected by the stellar merger.

%% file: appendix/companion_star_info.tex
\section{Companion stars: mass, radius and virial temperature.} \label{sec:companion_params}

\begin{figure}[!htbp]
   \centering
   \includegraphics[height=3in]{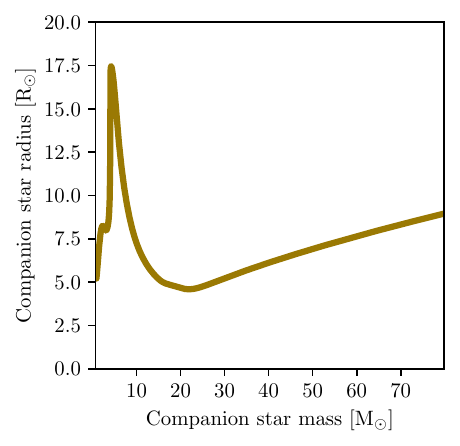} 
      \caption{Mass–radius relation for companion stars used in the collision simulations. The stellar radii are plotted as a function of companion mass, this relation is used to to determine the companion's minimum cross section as well as its virial temperature.
              }
         \label{fig:comp_mass_radius}
   \end{figure}

The companion star models were built with \texttt{MESA} trough cold gas accretion following directly the prescription by \cite{haemmerle2019stellar}, in the same manner as in \cite{ramirezga2025}. We extract the mass radius relationship from those stellar models as it is shown in Figure~\ref{fig:comp_mass_radius}. As well, the companion's virial temperature is computed as
\begin{equation}
    T_{\rm virial,comp} = \frac{G\ \mu\ m_{\rm H}\ m_{\rm comp}}{k_{\rm B}\ R_{\rm comp}}\ ,
\end{equation}
Where $\mu$ is the companion's mean molecular weight , $m_{\rm H}$ is the mass of the hydrogen atom and $k_{\rm B}$ is the Boltzmann constant. The values for $m_{\rm comp}$, $R_{\rm comp}$ and $T_{\rm virial,comp}$ for this work are shown in Table~\ref{tab:comp_params}.

\begin{table}[!htbp]
\caption{Stellar radius and virial temperature for the companion masses considered for this stellar merger simulations of this work.}             
\label{tab:comp_params}                
\centering                              
\begin{tabular}{c c c}                  
\hline\hline                            
$m_{\rm comp}$ [M$_\odot$] & $R_{\rm comp}$ [R$_\odot$] & $T_{\rm virial,comp}$ [$10^6$ K] \\ 
\hline                                 
10 & 7.25 & 19.13 \\
15 & 5.18 & 40.15 \\
20 & 4.66 & 59.51 \\
30 & 5.22 & 79.69 \\
40 & 6.11 & 90.78 \\
50 & 6.91 & 100.33 \\
60 & 7.64 & 108.90 \\
70 & 8.32 & 116.66 \\
\hline                                   
\end{tabular}
\end{table}

%% file: appendix/beta_values.tex
\section{On the remaining fraction of the deposited energy at the end of the inspiral} \label{sec:beta_values}

\begin{figure}[htbp!]
   \centering
    \includegraphics[height=3in]{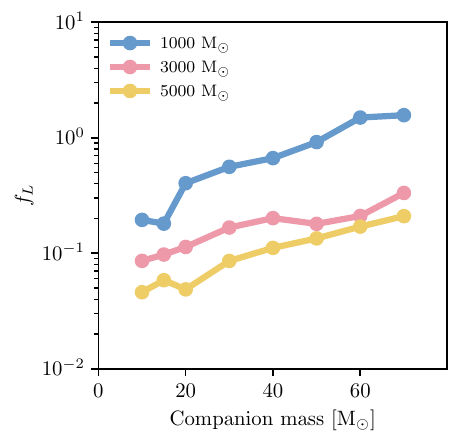} 
   \caption{Fractional change in luminosity, $f_L$, averaged from the inspiral start until the merger condition is reached, as function of companion mass. The different colors correspond to the three EMS masses adopted in this study, i.e. 1000, 2000 and 3000 \Msun.}
   \label{fig:fl_values}
\end{figure}

To estimate the amount of deposited orbital energy remaining at merger, the values of the luminosity change fraction $f_L$ are calculated following eq.~\ref{eq:lum_frac}. The resulting values for this parameter are shown in Figure~\ref{fig:fl_values} for all EMS mass and companion mass configurations. 

\begin{figure}[htbp!]
   \centering
    \includegraphics[height=3in]{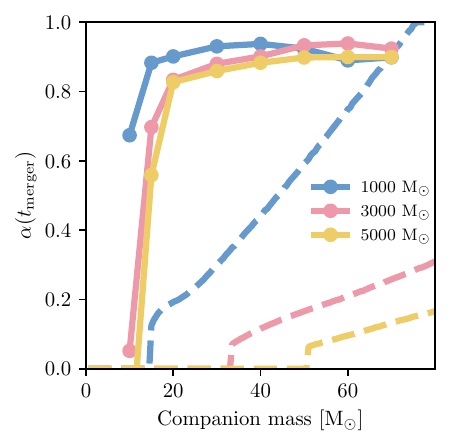} 
   \caption{Fraction of the deposited orbital energy remaining at merger, $\alpha(t_{\rm merger})$, as a function of companion mass. Solid-dotted lines show simulation results, while dashed lines show the analytical estimates given the assumptions by \cite{ramirezga2025}. Colors indicate the three EMS masses considered, i.e. 1000, 2000 and 3000 \Msun.}
   \label{fig:beta_values}
\end{figure}

Then, the fraction of orbital energy remaining at merger with respect to the total injected energy, $\alpha(t_{\rm merger})$, as defined in eq.~\ref{eq:beta_def}, is shown in Figure~\ref{fig:beta_values}. The solid-dotted lines correspond to the values of $\alpha$ obtained from our simulations, while the dashed lines represent the analytical estimates from \cite{ramirezga2025}, based on their assumptions for the fraction of deposited orbital energy available to drive mass loss. The different colors indicate the three EMS masses considered in this work. A direct comparison shows that the analytical prescription of \cite{ramirezga2025} systematically underestimates $\alpha$, and therefore underestimates the amount of deposited orbital energy available to drive mass loss.

Our simulations also reveal that collisions with low-mass companions yield significantly smaller values of $\alpha$. By rewriting eq.~\ref{eq:beta_def} as in eq.~\ref{eq:beta_def_timescale}, we identify two main factors responsible for this trend:
\begin{itemize}
    \item The smaller gravitational cross-section of low-mass companions leads to considerably longer merger timescales, $t_{\rm merger}$, compared to those for high-mass companions.
    \item Due to their lower mass, the total orbital energy deposited by such companions is expected to be smaller than that injected by high-mass ones, which in turn requires less time for the EMS to radiate away this energy.
\end{itemize}
These two effects combined reduce the fraction of deposited orbital energy remaining at merger for low-mass companions.

%% file: appendix/mass_loss_absolute.tex
\section{Mass-loss in absolute values} \label{sec:absolute_massloss}

\begin{figure*}
   \centering
    \includegraphics[height=2.8in]{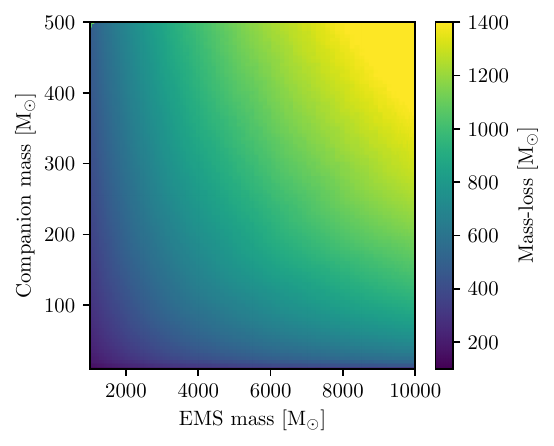} 
    \hspace{0.0in}
    \includegraphics[height=2.8in]{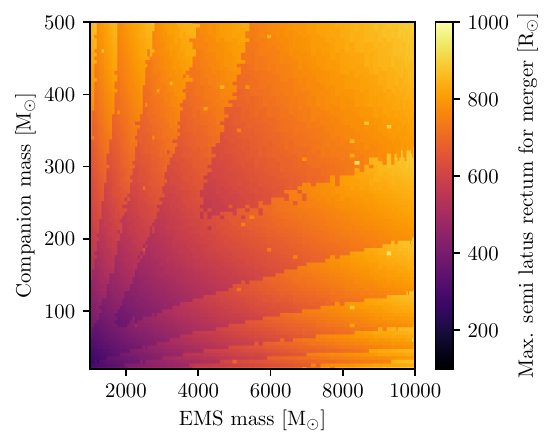} 
   \caption{Absolute estimates of merger-induced mass-loss and the latus rectum threshold between stellar mergers and scatterings, for orbits with initial eccentricity of 1.01. Considering EMS masses from $10^3$ to $10^4$~\Msun\ and companion masses from 5 to 500~\Msun. {Left plot:} maximum merger-induced mass-loss in M$_\odot$. {Right plot:} maximum initial latus rectum for a stellar merger in R$_\odot$.}
   \label{fig:max_massloss_lr_abs}
\end{figure*}

As in Figure~\ref{fig:max_massloss_lr_abs}, we show in Figure~\ref{fig:max_massloss_lr} the maximum mass-loss estimate and the latus rectum threshold for stellar mergers in absolute values, instead of the relative ones to the EMS mass and radius respectively. For further context in this figure see sec.~\ref{sec:across_ecc}.

